\documentclass{article}
\usepackage{spconf,graphicx, amssymb,amsmath,amsfonts,braket,multirow,arydshln,booktabs,array,subcaption}

\usepackage[
backend=biber,
style=ieee,
citestyle=numeric-comp,
maxbibnames=3,
maxcitenames=3,
doi=false,isbn=false,url=false,eprint=false
]{biblatex}

\defbibheading{bibliography}[\refname]{}

\newcommand{\refeq}[1]{Eq.~(\ref{eq:#1})}
\newcommand{\reftb}[1]{Table \ref{tab:#1}}
\newcommand{\reffig}[1]{Figure \ref{fig:#1}}

\title{LV-CTC: Non-autoregressive ASR with CTC and latent variable models}
\name{Yuya Fujita$^1$, Shinji Watanabe$^2$, Xuankai Chang$^2$, Takashi Maekaku$^1$}
\address{$^1$ LY Corporation, Tokyo, Japan, $^2$Carnegie Mellon University, PA, USA}
\copyrightnotice{979-8-3503-0689-7/23/\$31.00~\copyright2023 IEEE}
\begin{document}

\ninept
\maketitle
\begin{abstract}
\vspace{-0.95mm}
Non-autoregressive (NAR) models for automatic speech recognition (ASR) aim to achieve high accuracy and fast inference by simplifying the autoregressive (AR) generation process of conventional models. 
Connectionist temporal classification (CTC) is one of the key techniques used in NAR ASR models. 
In this paper, we propose a new model combining CTC and a latent variable model, which is one of the state-of-the-art models in the neural machine translation research field. 
A new neural network architecture and formulation specialized for ASR application are introduced.
In the proposed model, CTC alignment is assumed to be dependent on the latent variables that are expected to capture dependencies between tokens.
Experimental results on a 100 hours subset of Librispeech corpus showed the best recognition accuracy among CTC-based NAR models. 
On the TED-LIUM2 corpus, the best recognition accuracy is achieved including AR E2E models with faster inference speed.
\end{abstract}

\vspace{-1mm}
\setlength{\abovedisplayskip}{3pt} 
\setlength{\belowdisplayskip}{3pt}

\vspace{-1.6mm}
\section{Introduction}
\vspace{-0.2mm}
\label{sec:intro}
Automatic speech recognition (ASR) is a technology which has been widely used in the real world speech interface.
In most cases, faster inference with high recognition accuracy is preferred. 
For example, an ASR engine for the input method of a smartphone is required to return the recognition result as soon as possible after the end of an utterance for a better user experience. 
Another example is automatic captioning of user generated audios or videos whose length might be tens of thousands of hours in total, which demands huge computing resources to process all the contents in a realistic time.

One of the successful models of ASR is an autoregressive (AR) end-to-end (E2E) one \cite{AttentionNIPS2015, Graves2012, LAS2016, prabhavalkar2023endtoend}. 
The E2E model is comprised of a single neural network (NN).
It receives an acoustic feature sequence then generates a recognition hypothesis in a sequential manner by feeding back the previously generated token to the decoder.
Specifically, Transformer \cite{TransformerNIPS2017} architecture has been a fundamental building block of recently proposed E2E models \cite{Karita2019, karita2019comparative, dong2018speech}.  
It can achieve higher accuracy, but the sequential nature of the decoder limits the efficient use of the parallel computing capability of GPUs or ASICs specialized for NN computation.
By utilizing such a capability, aiming to not only achieve faster inference but also reduce power consumption, which should be strictly controlled in an on-device application, non-autoregressive (NAR) E2E model was proposed in the research field of neural machine translation (NMT) \cite{gu2018non, lee2018deterministic, Marjan2019}.
It eliminates or eases such sequential processes by generating multiple tokens in one iteration step.
It can efficiently use the parallel computing capability and achieve faster decoding at the expense of a drop in accuracy compared with AR models.
Some NAR E2E models have been proposed for ASR \cite{chan2020imputer, higuchi2021improvedMaskCTC, fujita20_interspeech, chi2021align, Chen2020Listen, fan2021cass, yu2021cif} and it is reported that they achieved competitive performance to AR models with faster inference speed under certain conditions \cite{higuchi2021comparative}.
The basic formulation of NAR models assume the offline decoding case however, they can be applied to the streaming case by using block-wise decoding \cite{wang21ba_interspeech, fujita21b_interspeech} or using them as the second pass refinement of a streaming model \cite{wang22delib_nar}.

In such NAR models for ASR, connectionist temporal classification (CTC) \cite{Graves2006CTC} and its variants \cite{lee2021intermediate, nozaki21_interspeech, higuchi2021improvedMaskCTC, higuchi2022hierarchical} are widely used. 
CTC has a monotonic alignment property which is thought to be reasonable for ASR because a sentence is read in left-to-right order.
Another major property of CTC is the assumption of conditional independence between tokens.
On the other hand, in the research field of NMT, latent variable models are applied as a way to relax the conditional independence assumption of NAR E2E models \cite{kaiser2018fast,shu2020latent}.
They introduce latent variables and assume the output token is dependent on the latent variable space.
It is expected that the latent variable captures dependencies between tokens and achieves competitive translation quality compared to AR models with faster inference speed.
Inspired by the success of latent variable models in NMT, we propose a NAR E2E ASR based on latent variable models. 
It is quite natural to apply it to ASR, however, to the best of the author's knowledge, there is no prior work of NAR E2E ASR based on latent variable models.

In this paper, a new model for ASR which combines CTC and latent variable models is proposed.
A new architecture of NN and its formulation are introduced.
The model can generate a hypothesis using the prior estimator network of the latent variable, which looks at only the acoustic feature sequence, by a single step.
It can also refine the hypothesis in an iterative way by feeding back the generated hypothesis to the posterior estimator network which looks at both the acoustic features and the token sequences.
The architecture of the NN and its formulation allow the proposed model to theoretically guarantee the performance of CTC and can improve its accuracy not only by iterative decoding but also by introducing additional techniques like intermediate CTC \cite{lee2021intermediate}.

Experiments are conducted on a 100 hours subset of the Librispeech \cite{Libri2015} and TED-LIUM2 \cite{tedlium2} corpora.
By intensive hyperparamter tuning and the combination of intermediate CTC \cite{lee2021intermediate} and self-distillation \cite{moriya20_interspeech} techniques,
the proposed model achieved the best accuracy among CTC-based NAR models on Librispeech.
On TED-LIUM2, the proposed model outperformed not only NAR models but also state-of-the-art AR models based on RNN-T and CTC/attention hybrid.

\section{Related Work}
\label{sec:related_work}
One of the unique properties of the proposed model is that the variational approximate posterior over latent variables is explicitly dependent on the output token sequence and the CTC alignment is assumed to be dependent on these latent variables.
None of the existing NAR E2E ASR using CTC \cite{chan2020imputer, higuchi2021improvedMaskCTC, fujita20_interspeech, chi2021align, fan2021cass, lee2018deterministic, chen21q_interspeech, nozaki21_interspeech} assumes such a dependency of CTC alignment in this latent space. 
For example, the combination of an insertion-based model and CTC proposed in \cite{fujita20_interspeech} explicitly assumes that the CTC alignment is dependent on a partial hypothesis, but not on latent variables.
Another unique aspect of the proposed model is that it can additionally employ techniques used in encoder-decoder architectures like a masked language model (MLM) \cite{Marjan2019}, glancing language model (GLM) \cite{qian2021glancing}, and self-distillation \cite{moriya20_interspeech}. 

\section{Methods}
\subsection{General formulation of E2E ASR and CTC}
E2E ASR utilizes a NN to model the following posterior distribution $p(C|X)$ over a token sequence $C=(c_n \in \mathcal{V} | n=1,\cdots,N )$, given a $d$-dimensional acoustic feature sequence $X=(\mathbf{x}_t \in \mathbb{R}^{d} | t=1,\cdots,T)$:
\begin{equation}
 \label{eq:basic_e2e}
 p(C|X) = \text{NN}(C, X;\theta).
\end{equation}
$N, T$ are the length of token and acoustic feature sequences, and $\mathcal{V}$ is a set of distinct tokens. $\theta $ is the parameters of the NN.
The difference between various E2E models is how to define the $p(C|X)$ and the NN architecture of $\text{NN}(C, X;\theta)$ in \refeq{basic_e2e}.
For example, attention-based encoder decoder (AED) \cite{AttentionNIPS2015, LAS2016} assumes left-to-right generation of a token sequence:
\vspace{-2mm}
\begin{equation}
\vspace{-1mm}
\label{eq:aed-1}
    p(C|X) =
    \prod _{n=1} ^{N} p(c_n | X, c_1, \cdots, c_{n-1}).
\end{equation} 
Then, the posterior of the $n$-th token, $p(c_n | X, c_1, \cdots, c_{n-1})$ in \refeq{aed-1}, is modeled by an encoder and decoder network through an attention mechanism.

CTC, which is one of the E2E models we focus on in this paper, introduces a latent alignment sequence and a mapping function $\mathcal{F}(\cdot)$. The mapping function $\mathcal{F}(\cdot)$ deletes the repetition of the same token and a special {\it blank} token.
Then, the posterior of a token sequence is defined as a summation over all the alignments that gives the same token sequence through the mapping function $\mathcal{F}(\cdot)$.

The alignment sequence $A$ is defined as a sequence of tokens with $\mathtt{\braket{b}}$, which is the special {\it blank} token, and the length is the same as the acoustic feature sequence:
\begin{equation}
    A=(a_{t} \in \mathcal{V} \cup \mathtt{\braket{b}}|t=1,\cdots,T).
\end{equation}
Then, the posterior over token sequence is defined as follows:
\begin{align}
\left\{
\label{eq:ctc-1}
\begin{alignedat}{4} 
    p(C|X) &= \sum  _{A \in \mathcal{F}^{\text{-1}}(C)} \prod _{t=1}^T p(a_t|X), \\
    p(a_t|X) &= \text{NN}(X;\theta).
    \end{alignedat} \right.
\end{align} 
The encoder layer of Transformer \cite{TransformerNIPS2017} and its variants \cite{gulati20_interspeech, kim2023branchformer} can be used as the $\text{NN}(\cdot)$ in \refeq{ctc-1}.
Conditional independence is assumed because in \refeq{ctc-1}, the posterior of the alignment depends only on the acoustic feature sequence, unlike \refeq{aed-1}, so the dependency between tokens is not taken into account.
 
\subsection{Latent variable models}
In latent variable models, the posterior in \refeq{basic_e2e} is assumed to be marginalized over $d^{\text{lat}}$-dimensional latent variable $Z= ( z_u \in \mathbb{R}^{d^{\text{lat}}} | u=1,\cdots, U )$ whose length is $U$:
\begin{equation}
 \label{eq:latent_model}
 p(C|X) = \int p(C|Z,X) p(Z|X)dZ.
\end{equation}
In general, the integral of \refeq{latent_model} is intractable, hence variational approximate posterior $q(\cdot)$ is introduced and the lower bound, which is also called evidence lower bound (ELBO), is maximized:
\begin{align}
     \mathcal{L}^{\text{ELBO}} &= \mathbb{E}_{Z \sim q } \left[ \log p^{\text{dec}}(C|Z,X)\right] \nonumber \\
     \label{eq:elbo}
     & ~~~~~~~~~~~~~~~~~~~~ - D^{\text{KL}}\left[q(Z|C,X) || p^{\text{prior}}(Z|X) \right],
\end{align}
where $D^{\text{KL}}(\cdot)$ is the Kullback-Leibler (KL) divergence. 
The three distributions, $p^{\text{dec}}(\cdot), q(\cdot)$, and $p^{\text{prior}}(\cdot)$ in \refeq{elbo} are modeled by NNs.
The ELBO can be maximized using the reparameterization trick \cite{Kingma2014} by assuming a Gaussian distribution over the latent variable $Z$.

In addition, depending on the type of $p^{\text{dec}}(\cdot)$, a length prediction module, which estimates the length $U$ of the latent variable $Z$, is introduced. 
For example, if AED is used as in \cite{shu2020latent}, the length prediction module is expected to estimate the length of the output token sequence.

At the inference stage, single step decoding can be performed by sampling $Z$ from the prior $p^{\text{prior}}(Z|X)$ and then feeding it to the decoder $p^{\text{dec}}(C|X,Z)$. 
Another way is to find a hypothesis which maximizes the ELBO in \refeq{elbo} using an algorithm with iteration.
One such algorithm is proposed in \cite{shu2020latent}, which does not require sampling or beam search.

\begin{figure*}[t]
\centering
\includegraphics[width=5.8in]{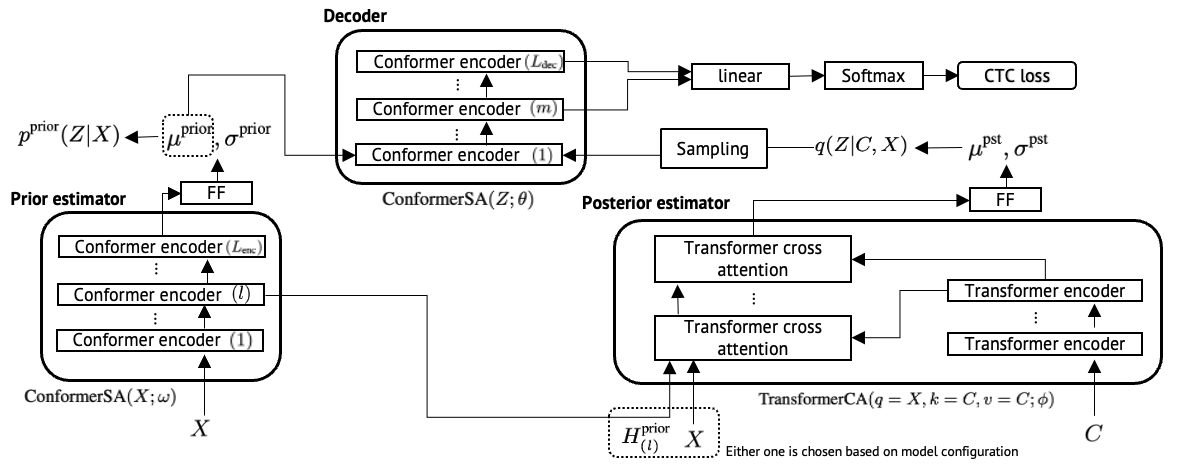}
\vspace{-4mm}
\caption{Architecture of the proposed model.}
\label{fig:structure}
\vspace{-6mm}
\end{figure*}

\subsection{Proposed CTC-based ASR using latent variable models}
\subsubsection{Basic architecture}
In the proposed method, CTC is used as the decoder\footnote{When CTC is used, the NN has only an encoder block but, by following the convention of latent variable models, we call the NN as decoder.}
$p^{\text{dec}}(\cdot)$ in \refeq{elbo} and the alignment posterior of the CTC in \refeq{ctc-1} is assumed to be dependent only on the latent variable $Z$:
\vspace{-2mm}
\begin{align}
\label{eq:ctc_dec}
    p^{\text{dec}}(C|Z,X) & \triangleq p(C|Z) 
    = \sum  _{A \in \mathcal{F}^{\text{-1}}(C)} \prod _{t=1}^T p(a_t|Z). 
\end{align} 
\vspace{-1mm}
The reasons for employing CTC are twofold: (1) it does not require a length prediction module and (2) the monotonic alignment property is thought to be reasonable for ASR.

Then, a Transformer or Conformer architecture is used as the NN of each component of \refeq{elbo} and \refeq{ctc_dec}.
The alignment posterior $p(a_t|Z)$ in \refeq{ctc_dec} is modeled by Conformer self-attention layers, which are depicted as ``Decoder" on the top of \reffig{structure}:
\begin{flalign}
\label{eq:ctc_decoder}
 & \left\{ 
\begin{alignedat}{4} 
   p(a_t|Z) 
     &= \text{Softmax}\left( \text{Linear}(H^{\text{dec}}; \theta^{\text{out}}) \right), \\
   H^{\text{dec}} &= \text{ConformerSA}(Z; \theta), \\
  Z & \sim q(Z|C,X).
\end{alignedat} \right. &
\end{flalign}
The variational approximate posterior $q(\cdot)$ in \refeq{elbo} is modeled by the cross-attention layers of Transformer, which is depicted as ``posterior estimator" at the bottom right of \reffig{structure}:
\begin{flalign}
 & \left\{ 
\begin{alignedat}{4} 
\label{eq:pst_enc}
   q(Z|C,X) & = \mathcal{N}(\mu^{\text{pst}}, \sigma^{\text{pst}}),\\
   H^{\text{pst}} &= \text{TransformerCA}(q=X, k=C,v=C; \phi), \\
    \mu^{\text{pst}} &= \text{FF}^{\text{pst}}(H^{\text{pst}};\phi^{\text{m}}), \\
    \sigma^{\text{pst}} &= \text{FF}^{\text{pst}}(H^{\text{pst}}; \phi^{\text{s}}).
\end{alignedat} \right. &
\end{flalign} 
The prior $p^{\text{prior}}(\cdot)$ in \refeq{elbo} is modeled by other self-attention layers of Conformer, which is depicted as ``prior estimator" in \reffig{structure}:
\begin{flalign}
 & \left\{ 
\begin{alignedat}{4} 
\label{eq:prior_enc}
    p^{\text{prior}}(Z|X) & = \mathcal{N}(\mu^{\text{prior}}, \sigma^{\text{prior}}),\\
    H^{\text{prior}} &= \text{ConformerSA}(X; \omega), \\
    \mu^{\text{prior}} &= \text{FF}^{\text{prior}}(H^{\text{prior}};\omega^{\text{m}}), \\
    \sigma^{\text{prior}} &= \text{FF}^{\text{prior}}(H^{\text{prior}}; \omega^{\text{s}}). \\
\end{alignedat} \right.  &
\end{flalign}
The meanings of each component are:
\begin{itemize}
\vspace{-1mm}
    \setlength{\leftskip}{3.0cm}
    \setlength{\itemsep}{-0.8mm}
    \item[$\text{ConformerSA}(\cdot; \psi)$ :] Self-attention layers (encoder block) of Conformer
    \item[$\text{TransformerCA}(q,k,v; \psi)$ :] Cross-attention layers (decoder block) of Transformer where $q, k, v$ are query, key, and value, respectively
    \item[$\text{FF}(\cdot; \psi )$ :] Feed forward layer
    \item[$\text{Linear}(\cdot; \psi)$ :] Linear layer
    \vspace{-1mm}
\end{itemize}
where $\psi$ is the parameter of the component.
The whole architecture is depicted in \reffig{structure}.

By employing this architecture, the latent variables $Z$ are expected to capture token dependencies because the minimization of KL divergence between the posterior $q(Z|C,X)$, which looks at both entire acoustic feature $X$ and token sequences $C$, and the prior $p^{\text{prior}}(Z|X)$ reflects the token dependency inside $C$.

\subsubsection{Additional techniques}
In addition to the basic architecture introduced in the previous section, the following techniques are introduced which are expected to achieve higher accuracy. 

\vspace{-6mm}
\paragraph*{Compatibility with vanilla CTC}
As CTC is a strong baseline, we make the proposed entire network to be compatible with vanilla CTC, so that the training would be stable and guarantee the accuracy of vanilla CTC.
This is realized by feeding the mean value calculated by the prior estimator $\mu^{\text{prior}}$ in \refeq{prior_enc} into the decoder's self-attention layers $\text{ConformerSA}(\cdot; \theta)$ in \refeq{ctc_decoder} and obtain the alignment posterior similar to \refeq{ctc_dec}:
\begin{align}
\label{eq:cp_ctc}
& \left\{ 
\begin{alignedat}{4} 
    p(a_t|Z=\mu^{\text{prior}}) 
     &= \text{Softmax}\left( \text{Linear}(H^{\text{dec}}_{\text{prior}}; \theta^{\text{out}}) \right), \\
    H^{\text{dec}}_{\text{prior}} &= \text{ConformerSA}(Z=\mu^{\text{prior}}; \theta).
\end{alignedat} \right.
\end{align}
Then, the CTC loss is computed using the alignment posterior and is jointly trained with ELBO of \refeq{elbo},
\begin{equation}
    \mathcal{L}^{\text{ctc}}_{\text{cp}} = \log \sum  _{A \in \mathcal{F}^{\text{-1}}(C)} \prod _{t=1}^T p(a_t|Z=\mu^{\text{prior}}).
\end{equation}

\vspace{-8mm}
\paragraph*{Sharing encoder layer between prior and posterior estimator}
In the preliminary experiment, it is observed that the posterior estimator's source-target attention $\text{TransformerCA}(\cdot)$ in \refeq{pst_enc} tends to be unstable. 
It might be because the acoustic feature is not well transformed to fit to the source-target attention where similarity between transformed features from a token sequence is calculated.
Therefore, the intermediate output of prior estimator's encoder layer $\text{ConformerSA}(\cdot;\omega)$ in \refeq{prior_enc} is fed into the $\text{TransformerCA}(\cdot)$ in \refeq{pst_enc}:
\begin{equation}
    \label{eq:enc_share}
    H^{\text{pst}} = \text{TransformerCA}(q=H^{\text{prior}}_{(l)}, k=C,v=C; \phi), 
\end{equation}
where $H^{\text{prior}}_{(l)}$ is the $l$-th layer's output of $\text{ConformerSA}(\cdot;\omega)$ in \refeq{prior_enc}.
In \reffig{structure}, this sharing is depicted as the arrow from the prior estimator on the left side to the posterior estimator on the right side.

\vspace{-6mm}
\paragraph*{Intermediate CTC loss}
Intermediate CTC \cite{lee2021intermediate} is a technique adding CTC losses at the intermediate layers of the encoder block and accuracy improvement is reported. 
We employed the technique in our proposed architecture by adding an extra CTC loss at an intermediate layer of decoder's $\text{ConformerSA}(\cdot; \theta)$ in \refeq{cp_ctc} and \refeq{ctc_decoder}:
\begin{align}
\label{eq:interctc}
& \left\{ 
\begin{alignedat}{4} 
    p^{\text{prior}}(a_t|H^{\text{dec}}_{\text{prior}, (m)}) &= 
    \text{Softmax}\left( \text{Linear}(H^{\text{dec}}_{\text{prior}, (m)}; \theta^{\text{out}}) \right), \\
    p^{\text{pst}}(a_t|H^{\text{dec}}_{(m)}) &= 
    \text{Softmax}\left( \text{Linear}(H^{\text{dec}}_{(m)}; \theta^{\text{out}}) \right),
\end{alignedat} \right.
\end{align}
where $H^{\text{dec}}_{\text{prior}, (m)}$ is the output of the $m$-layer of $\text{ConformerSA}(\cdot; \theta)$ in \refeq{cp_ctc} and $H^{\text{dec}}_{(m)}$ is the output  of $m$-th-layer of $\text{ConformerSA}(\cdot;\theta)$ of \refeq{ctc_decoder}.
Note that the parameters of the Linear layer $\theta^{\text{out}}$ is shared.
Then, the CTC loss is computed using the alignment posterior and they are jointly trained with ELBO of \refeq{elbo}:
\begin{align}
\label{eq:interctc_loss}
& \left\{ 
\begin{alignedat}{4} 
\mathcal{L}^{\text{ictc}}_{\text{prior}} &= \log \sum  _{A \in \mathcal{F}^{\text{-1}}(C)} \prod _{t=1}^T p(a_t|H^{\text{dec}}_{\text{prior}, (m)}), \\
\mathcal{L}^{\text{ictc}}_{\text{pst}} &= \log \sum  _{A \in \mathcal{F}^{\text{-1}}(C)} \prod _{t=1}^T p(a_t|H^{\text{dec}}_{(m)}).
\end{alignedat} \right.
\end{align}
It is also expected that the instability of source-target attention of $\text{TransformerCA}(\cdot)$ in \refeq{pst_enc} mentioned before can be mitigated.

\vspace{-4mm}
\paragraph*{Self-distillation}
Self-distillation (SD) \cite{moriya20_interspeech} is a technique that was originally proposed to perform distillation in AED model by assuming the output of the decoder as teacher and the encoder is the student during training from scratch \cite{moriya20_interspeech}.
In our proposed model architecture, the alignment posterior of \refeq{ctc_decoder} which is computed by the latent variable sampled from the posterior in \refeq{pst_enc} can be viewed as the teacher because it looks at the entire token sequence.
Then, the alignment posterior of \refeq{cp_ctc} becomes the student and the KL-divergence between them are added to the ELBO of \refeq{elbo}:
\begin{equation}
    \mathcal{L}^{\text{SD}} = - \sum _t D^{\text{KL}}\left( 
        p(a_t|Z=\mu^{\text{prior}}) || p(a_t|Z) 
        \right).
\end{equation}
\vspace{-2.5mm}

\subsubsection{The loss function}
The loss function of the proposed model is the summation of all the losses defined so far. By introducing coefficients, $\alpha_{(\cdot)}$, to adjust the dynamic range of each loss, the loss function is defined as follows:
\begin{align} 
 \label{eq:loss}
    \mathcal{L} &= \underbrace{\alpha_{\text{dec}} \mathbb{E}_{Z \sim q } \left[ \log p^{\text{dec}}(C|X,Z)\right] -
     \alpha_{\text{KL}} D^{\text{KL}}\left[q(Z|C,X) || p^{\text{prior}}(Z|X) \right]}_{\mathcal{L}^{\text{ELBO}}}\nonumber \\
    &
    + \alpha_{\text{cp}}\mathcal{L}^{\text{ctc}}_{\text{cp}}
    + \alpha_{\text{ic1}}\mathcal{L}^{\text{ictc}}_{\text{prior}}
    + \alpha_{\text{ic2}} \mathcal{L}^{\text{ictc}}_{\text{pst}}
    + \alpha_{\text{SD}} \mathcal{L}^{\text{SD}}.
\end{align}

\section{Experiments}
\begin{table}[t]
  \caption{WERs of ``dev" sets of LS-100 by changing number of layers of prior estimator $L_{\text{enc}}$, the decoder $L_{\text{dec}}$, and the index of encoder sharing $l$ in \refeq{enc_share}. Results without iteration (Greedy) and 3 iterations are shown.}
  \label{tab:res1}
   \vspace{-7mm}
 \begin{center}
    \renewcommand{\arraystretch}{0.77}
    \begin{tabular}{ccc|rr|rr|}
    &  &  &  \multicolumn{2}{c|}{\textbf{Greedy}} & \multicolumn{2}{|c|}{\textbf{3 iterations}} \\   
    $L_{\text{enc}}$ & $L_{\text{dec}}$ & $l$ & clean & other & clean & other \\
    \midrule   
3	& 12 &	- &	7.5 &	21.0 &	8.2 & 	23.4 \\
6 &	9 & &		7.5 &	21.4 &	8.4 &	23.7 \\
9 &	6 &	 &	7.9 &	22.3 &	7.9 &	22.1 \\
\midrule
3 &	12 &	1 &	7.3 &	21.0 &	7.1 &	19.9 \\
 & &	2 &	7.2 &	21.4 &	6.6 &	19.1 \\
6 &	9 &	2 & 8.2 &	22.8 &	88.1 & 	90.3 \\ 
9 &	6 &	3 &	7.7 &	22.1 & 7.2 &	20.3 
    \end{tabular}
 \end{center}
 \vspace{-9mm}
\end{table}

Two corpora, Librispeech \cite{Libri2015} and TEDLIUM2 (TED2) \cite{tedlium2} are used to evaluate the proposed method.
First, basic experiments of investigating the detail of the NN architecture are performed using a 100 hours subset of Librispeech (LS-100).
Then, based on the hyperparameters chosen by the LS-100 results, TED2 is evaluated.
Finally, comparison to some existing AR/NAR models are shown.

\subsection{Basic experiments and analysis on LS-100}
\subsubsection{Setup}
First, speed perturbation with scaling factor of 0.9 and 1.1 is applied and the perturbed data are added to the original training data.
The acoustic feature is 80 dimensional log Mel-filterbank. 
SpecAugment \cite{Park2019} is applied to the acoustic feature sequence whose parameters are set identical to the recipe of ESPnet \cite{Watanabe2018Espnet}.
For the output token, byte pair encoding (BPE) is applied with the vocabulary size of 100. 
Before it is fed into the $\text{TransformerCA}(\cdot)$ in \refeq{pst_enc}, an embedding layer which converts token ids into a real-valued vector is applied. 
Then, SpecAugment with only time masking is applied. 
The number of masks are set as 10\% of the length of the output token sequence. 

The network architecture of the proposed method is as follows.
The acoustic feature sequence is down-sampled to 1/4 of the original rate by using 2 layers of convolutional neural network (CNN) whose channels, stride, and kernel size are 256, 2, and 3, respectively.
Then, the output of the CNN is fed into $\text{ConformerSA}(\cdot)$ and $\text{TransformerCA}(\cdot)$ in Eq.~(\ref{eq:pst_enc})-(\ref{eq:prior_enc}).
For the attention modules of $\text{ConformerSA}(\cdot)$ and $\text{TransformerCA}(\cdot)$ in Eq.~(\ref{eq:ctc_decoder})-(\ref{eq:prior_enc}), relative positional encoding is used \cite{Dai2019}.
The dimension of the attention is set as 256 and the number of head is 4.
The length of kernel, number of hidden units of FF module, and the activation function of $\text{ConformerSA}(\cdot)$ in Eq.~(\ref{eq:ctc_decoder}), (\ref{eq:prior_enc}) are 15, 1024, and swish, respectively. 
The parameters of $\text{TransformerCA}(\cdot)$ in \refeq{pst_enc} are the same as $\text{ConformerSA}(\cdot)$ except that there is no CNN module in it. 
The number of hidden units of the $\text{FF}(\cdot)$ module in Eq.~(\ref{eq:pst_enc})-(\ref{eq:prior_enc}) is 1024 and the activation function is hyperbolic tangent.
For all components, the dropout rate is set as 0.1.

The training is run for 50 epochs using 4 Tesla V100 GPUs. 
For the optimizer and scheduler, Adam \cite{kingma2015adam} with $\beta_1 = 0.90, \beta_2 = 0.98$ and Noam scheduling \cite{TransformerNIPS2017} with warmup step of 15,000, are used. 
The peak learning rate is set as 0.002. The weight decay is 0.00001.
The decoding is performed using the averaged model over the top 10 validation scores. 

First, we show a basic experiment of changing the number of layers of $\text{ConformerSA}(\cdot)$ in Eq.~(\ref{eq:ctc_decoder}),(\ref{eq:prior_enc}) and the index of the layer shared between prior and posterior network, i.e., $l$ in \refeq{enc_share}\footnote{There are many other hyperparameters in the proposed model but, according to preliminary experiments, these parameters are crucial to obtain reasonable accuracy so the results are shown in the paper.}.
Note that the following parameters are fixed based on a preliminary experiment or intuition.
Number of layers of $\text{TransformerCA}(\cdot)$ in Eq.~(\ref{eq:pst_enc}) is set to 2 and $ \alpha_{\text{dec}},\alpha_{\text{KL}}$, and $\alpha_{\text{cp}}$ in Eq.~(\ref{eq:loss}) are set to 0.09, 0.1, and 0.81, respectively by intuition
\footnote{
It looks $\alpha_{\cdot}$ are too specific but it comes from the difference between actual implementation and the formulation of \refeq{loss}. In the experiments, actual values for each term were 0.1. 
}.
The dimension of the latent variable is 64.
In addition, when the KL-divergence of \refeq{loss} is smaller than $b$ in a training batch, $\alpha_{\text{KL}}$ is set to 0. $b=0.5$ is used. The last two parameters are fixed by preliminary experiments.
We plan to make the configuration files to be publicly available upon publication of the paper.

\begin{figure}[t]
      \begin{minipage}[t]{1.0\hsize}
        \centering
        \includegraphics[keepaspectratio, scale=0.11]{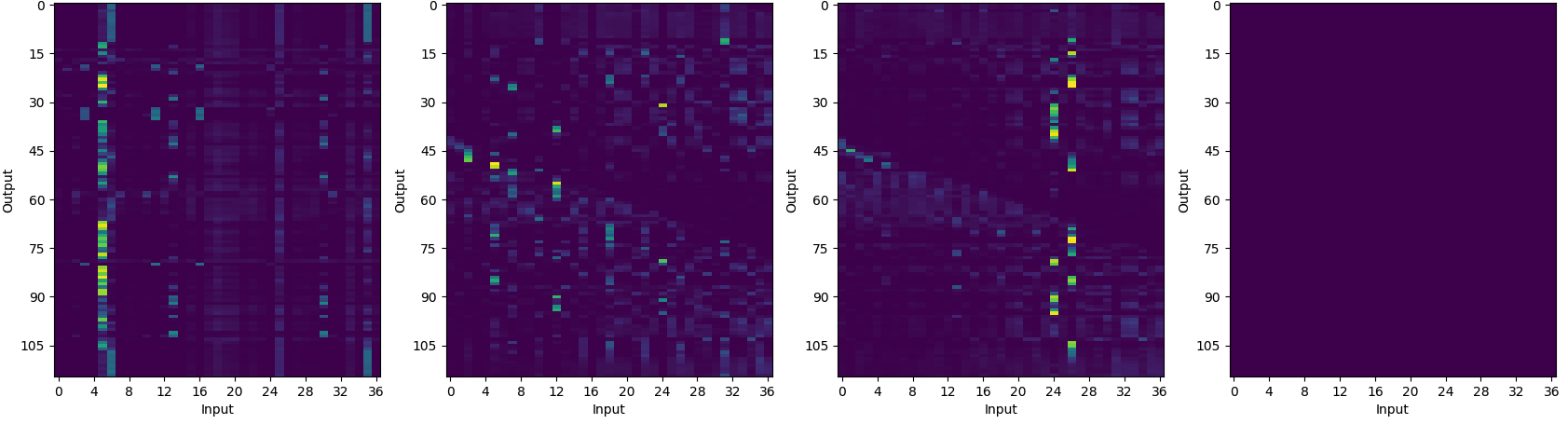}
        \vspace{-2mm}
        \subcaption{$L_{\text{enc}}=3, L_{\text{dec}}=12$}
        \label{a1}
      \end{minipage} \\
       \begin{minipage}[t]{1.0\hsize}
        \centering
        \includegraphics[keepaspectratio, scale=0.11]{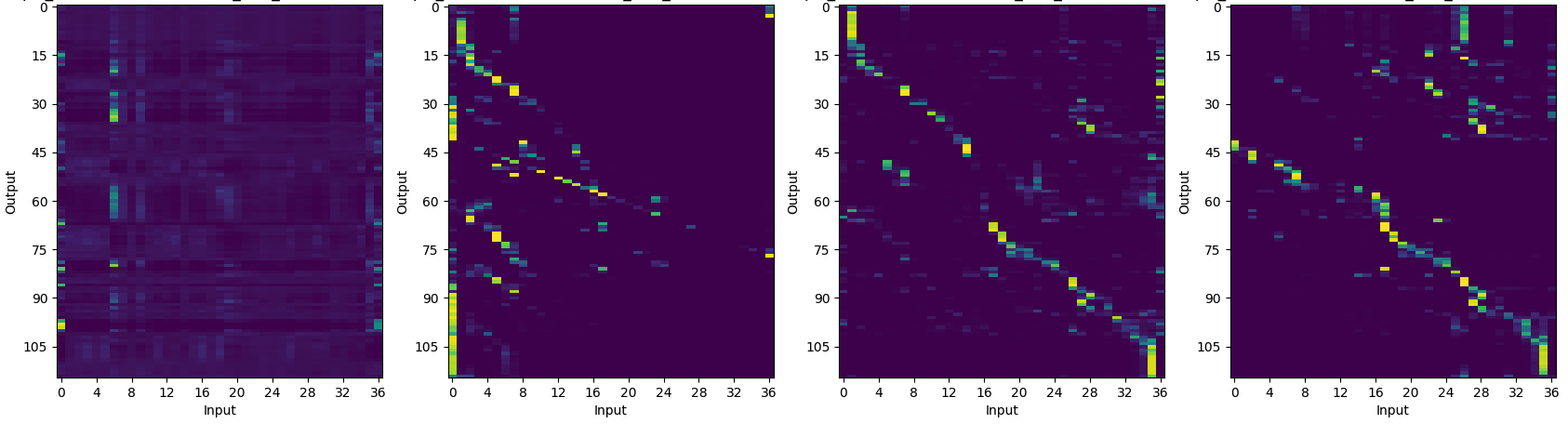}
        \vspace{-2mm}
        \subcaption{$L_{\text{enc}}=6, L_{\text{dec}}=9$}
        \label{a1}
      \end{minipage} \\
        \begin{minipage}[t]{1.0\hsize}
        \centering
        \includegraphics[keepaspectratio, scale=0.11]{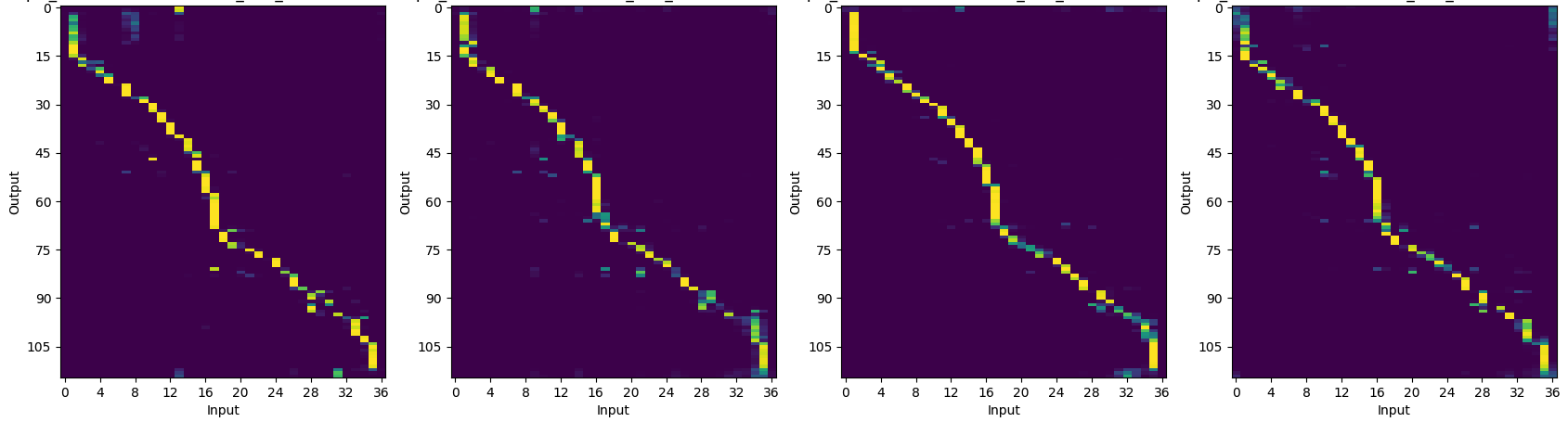}
        \vspace{-2mm}
        \subcaption{$L_{\text{enc}}=9, L_{\text{dec}}=6$}
        \label{a1}
      \end{minipage} \\
        \begin{minipage}[t]{1.0\hsize}
        \centering
        \includegraphics[keepaspectratio, scale=0.11]{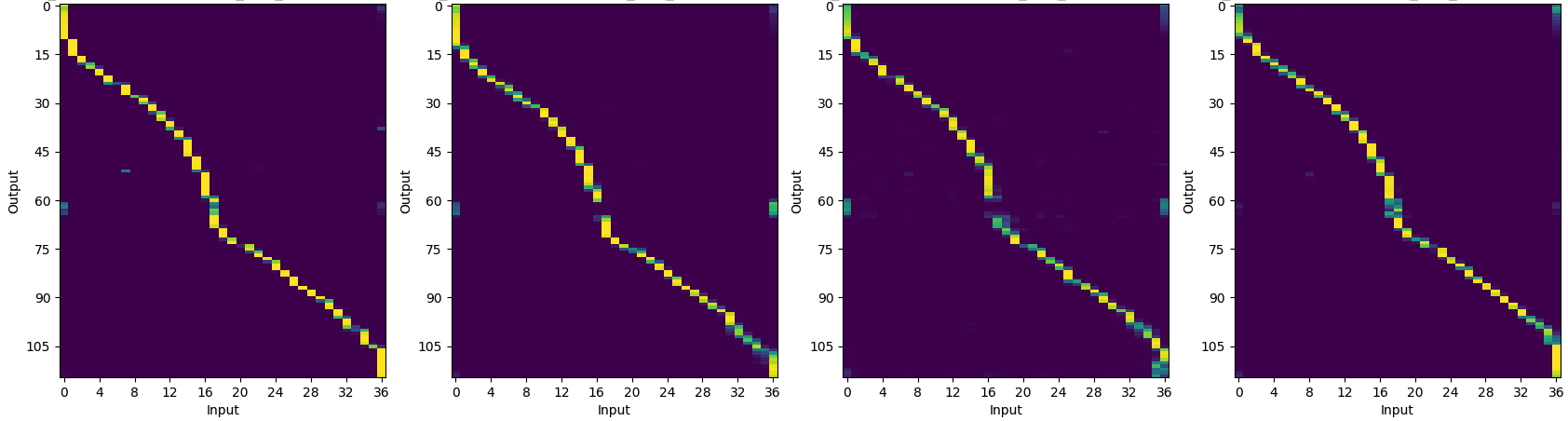}
        \vspace{-2mm}
        \subcaption{$L_{\text{enc}}=3, L_{\text{dec}}=12, l=2$}
        \label{a2}
      \end{minipage} 
       \vspace{-4mm}
       \caption{Visualization of the attention weight of the cross-attention of posterior estimator $\text{TransformerCA}(\cdot)$ in \refeq{pst_enc}. An utterance from training set of Librispeech is chosen. The horizontal axis is the index of token sequence and the vertical axis is the index of acoustic feature sequence.}
      \label{fig:att}
      \vspace{-6mm}
  \end{figure}

\begin{table*}[t]
  \caption{Comparison of WERs on LS-100 and TED2. The RTFs are measured on ``dev-other" of Librispeech. ``Speedup" is the improvement of RTF from Conformer-T decoded with beam size of 10.  }
  \label{tab:all_results}
 \begin{center}
 \vspace{-7mm}
 \renewcommand{\arraystretch}{0.81}
    \begin{tabular}{lc|cc|rrrr|rr|}
    \toprule
     \multirow{3}{*}{\textbf{Model}}     &   \multirow{3}{12mm}{\textbf{Beam or  \#Iter.}}  & \multirow{3}{*}{\textbf{RTF}} & \multirow{3}{*}{\textbf{Speedup}}       & \multicolumn{4}{c|}{\textbf{Librispeech}} & \multicolumn{2}{c|}{\multirow{2}{*}{\textbf{TEDLIUM2}}} \\    
          &        & &        & \multicolumn{2}{c}{\textbf{dev}} & \multicolumn{2}{c|}{\textbf{test}} \\ 
     &  &  &  & clean & other & clean & other & dev & test \\
    \midrule
AED & 1 & 0.299 & 0.97 &	6.7 &	18.1 &	7.1 &	18.2 &	7.7 &	8.4 \\ 
 + Beam search   & 10 & 0.797 & 0.37 &	6.1 &	17.9 &	6.4 & 17.9 & \bf 7.3 & \bf 7.8 \\
Conformer-T & 1 & 0.057 & 5.11 &	6.3 &	17.6 &	6.7 &	17.5 &	8.1 &	8.1 \\
 + Beam search    & 10 & 0.291 &	1.00 &	{\bf 5.7} &	{\bf 16.8} & \bf 6.2 &	\bf 16.8 &	8.0 &  8.0 \\
 \midrule
 CTC & 1 &  0.044 & 6.61 &	6.8 &	19.7 & 6.9 & 19.8 &	8.1 & 7.8 \\ 
 Intermediate CTC & 1 &  0.045 &	 6.47 & 6.1 & 18.8 & 6.8 & 18.9 & 7.2 & 7.5 \\ 
 SC-CTC & 1 &  0.047 & 6.19 &	6.1 & 18.9 & 6.4 &	19.1 &	7.3 & 7.7 \\ 
 LV-CTC (proposed) & 1 & 0.043 & 6.77 & 6.4 & 18.5 & 6.5 & 18.9 & 7.4 & 7.6  \\
 + Iteration & 3 & 0.084 & 3.46 & 6.0 & 16.8 & 6.1 & 17.2 & \bf 7.0 & 7.2  \\
 + SD & 1 & 0.040 & 7.28 & 6.2 & 18.3 & 6.5 & 18.5 & 7.5 & 7.5  \\
 ~~ + Iteration & 3 & 0.086 & 3.38 & \bf 5.9 & \bf 16.7 & \bf 6.1 & \bf 17.1 & 7.1 & \bf 7.1  \\
KERMIT \cite{fujita20_interspeech}    &  5 &  - &	- &	 6.4 & 17.9 & 6.6 & 18.2 &	8.6 & 8.0 \\
Improved Mask-CTC \cite{higuchi2021improvedMaskCTC, higuchi2021comparative} &	5 & - &	- &	7.0 &	19.8 &	7.3 &	20.2 &	8.8 &	8.3\\
SC-CTC \cite{nozaki21_interspeech, higuchi2021comparative} & 1 & - & - &		6.6 &	19.4 &	6.9 &	19.7 &	8.7 & 8.0 \\
HC-CTC \cite{higuchi2022hierarchical} & 1 & - & - & 		6.9 &  17.1 &	7.1 & 17.8 &	8.0 & 7.6 		       \\
\bottomrule
    \end{tabular}
 \end{center}
 \vspace{-10mm}
\end{table*}

\subsubsection{Results}
The number of layers $L_{\text{enc}}$ of $\text{ConformerSA}(\cdot;\omega)$ in Eq.~(\ref{eq:prior_enc}), the number of layers $L_{\text{dec}}$ of $\text{ConformerSA}(\cdot;\theta)$ in Eq.~(\ref{eq:ctc_decoder}), and $l$ in \refeq{enc_share} are searched.
The results are shown in \reftb{res1}.

From the upper side of the table, it can be seen that when $L_{\text{enc}}$ is small, i.e., the shallower the prior estimator is, better accuracy is obtained except for the case when $L_{\text{enc}} =9, L_{\text{dec}}=6$ and iterative decoding is performed.
In this case, the attention patterns of the cross attention of $\text{TransformerCA}(\cdot)$ of \refeq{pst_enc} look more monotonic than the other cases, as visualized in \reffig{att}. 
This might lead to slight improvements by iterative decoding.

From the bottom side of the table, the sharing of the encoder in the way of \refeq{enc_share} is effective except the case when $L_{\text{enc}}=6, L_{\text{dec}}=9$.
In this case, the iterative decoding fails almost completely. 
On the other hand, by setting $L_{\text{enc}}=3$, $L_{\text{dec}}=12$, and $l=2$, the best WER is achieved.
With this setting, the attention patterns of the cross attention of $\text{TransformerCA}(\cdot)$ of \refeq{pst_enc} look more monotonic as visualized in \reffig{att}d. 
According to these results, the monotonic property of the cross attention is important for obtaining better WERs with iterative decoding.
In addition, the proposed model is quite sensitive to the parameters of $L_{\text{enc}}, L_{\text{dec}}$, and $l$.

\subsection{Comparison to baseline models}
In addition to the best configuration of the previous section, intermediate CTC and self-distillation are introduced. 
The index of intermediate layer $m$ in \refeq{interctc} is set to 4.
When only the intermediate CTC is applied, the coefficients in \refeq{loss} are $(\alpha_{\text{dec}}, \alpha_{\text{KL}}, \alpha_{\text{cp}}, \alpha_{\text{ic1}}, \alpha_{\text{ic2}})=$(0.081, 0.1, 0.729, 0.009, 0.081).
When both intermediate CTC and self-distillation are applied, the coefficients are $(\alpha_{\text{dec}}, \alpha_{\text{KL}}, \alpha_{\text{cp}}, \alpha_{\text{ic1}}, \alpha_{\text{ic2}}, \alpha_{\text{SD}})=$(0.073, 0.1, 0.656, 0.008, 0.073, 0.090).
Then, the training run for 100 epochs.
Four models, Conformer-based AED \cite{guo2021}, Conformer-Transducer (Conformer-T) \cite{boyer2021study}, and Conformer-based CTC including intermediate CTC \cite{lee2021intermediate} and Self-conditioned CTC (SC-CTC) \cite{nozaki21_interspeech}, are reproduced and evaluated as baseline.

Conformer-based AED has 12 layers of Conformer encoder and 6 layers of Transformer decoder. The parameters are the same as the proposed model except that the attention module of the Transformer uses absolute positional embedding. Conformer-based CTC models have 18 layers of Conformer encoder layers whose parameters are the same as the proposed model. For intermediate CTC and SC-CTC, the 6-th and 12-th layers are used as intermediate layers. 
The Conformer-T has 17 layers of Conformer encoder whose parameters are the same as the proposed model and 1 layer of LSTM decoder whose hidden unit size is 420. The dimension of the joint network is 320. 
CTC loss is added to the encoder network during training with the weight of 0.3 for AED. 

These models are trained for 100 epochs using the same GPUs as the proposed model.
The settings of optimizer and scheduler are the same as the proposed model for all the baseline models. 
The decoding is performed using the averaged model over top 10 validation scores. 
For the AED model, joint CTC decoding \cite{Shinji2017hybrid} is used with the CTC weight of 0.3. Language model is not used for any of the models.
RTF is measured on the dev-other set of Librispeech on an Intel(R) Xeon(R) Gold 6138 CPU @ 2.00GHz CPU using 4 threads for NN inference. 

The results are shown in \reftb{all_results}. 
In addition to the reproduced baseline models, WERs of some related works are also shown.
When compared between baseline CTC models and the proposed LV-CTC without iteration, WERs of LV-CTC with self-distillation (SD) \footnote{In this experiment, the dropout rate of posterior encoder is increased to 0.2 because it performed better when trained until 100 epochs.} are better than CTC models on ``other" sets and are competitive on ``clean" sets.
From this result, the proposed architecture can at least maintain the performance of baseline CTC models.
The LV-CTC with iteration performs the best among all the NAR models shown in the table at the expense of around 2 times increase in RTFs compared with baseline CTC models.
It outperforms AED model even with beam search at smaller RTF.
When compared with Conformer-T, the proposed method achieved better WERs than greedy decoding results at the expense of the RTF increased to 1.5 times.
With beam search, they are competitive or the proposed method is slightly worse but the RTF of the proposed method is reduced to 0.3 times.

\subsection{Results on TED2}
Based on the best hyperparameter of LS-100 experiments, evaluation on TED2 is performed.
Most of the training configurations are the same except the number of epochs is 100. 
The difference from the best hyperparameter on LS-100 are as follows:
\begin{itemize}
\vspace{-4.5pt}
\setlength{\itemsep}{-0.9mm}
\item Number of Transformer encoders in $\text{TransformerCA}(\cdot)$ in \refeq{pst_enc} is increased from 2 to 3 because in the preliminary experiments $\text{TransformerCA}(\cdot)$ is diverged.
\item Increased dropout rate and the number of masks of SpecAugment of token embedding because there is a possibility of over-fitting.
\item Decreased self-distillation weight $\alpha_{\text{SD}}$ to 0.001 because when the weight is larger than it, WER degraded.
\vspace{-4.5pt}
\end{itemize}
The results are shown in \reftb{all_results}.
When compared between baseline CTC models and the proposed LV-CTC without iteration, WERs of baseline models are better than LV-CTC.
By using iterative decoding, the proposed LV-CTC has the best WER among all the models shown in the table.
Overall, the trend of WER on TED2 is different from that of the LS-100 case. 
Self-distillation improved test set but degraded dev set.
Even AR models' WERs are sometimes worse than baseline CTC models.
The speaking style is different between these two corpora: LS-100 is read speech and TED2 is presentation talks.
This might be the reason for the different trends but further investigation is left as future work.

\section{Conclusion}
In this paper, we proposed a new ASR model by combining CTC and latent variable models. 
By introducing a new architecture of NN and its formulation, the proposed model gave at least the same performance of vanilla CTC. 
In addition, iterative decoding could refine the hypothesis and achieved higher accuracy at the expense of increased inference speed. 
Experiments are conducted on a 100 hours subset of Librispeech and TED-LIUM2 corpora. 
On the Librispeech corpus, the proposed model achieved the best WER compared with CTC-based NAR models.
On the TED-LIUM2 corpus, the proposed model outperformed NAR and AR models.
Investigation of the different trends in WER between the two corpora is left as future work.

\section{Acknowledgement}
We would like to thank Jaesong Lee for introduction of his preliminary trial on latent variable models for ASR.

%
\vfill
\pagebreak


\section{References}
{
\renewcommand{\baselinestretch}{0.5}
\printbibliography
}

\end{document}